\documentclass[12pt]{iopart}
\usepackage{amssymb}

\usepackage{graphicx}
\usepackage{cite}

\newcommand{\beq}{\begin{eqnarray}}
\newcommand{\eeq}{\end{eqnarray}}

\begin{document}

\title{$\mathcal{PT}$-symmetric strings}
\author{Paolo Amore\dag, Francisco M. Fern\'andez\ddag, Javier Garcia\ddag, German
Gutierrez\dag}

\address{\dag\ Facultad de Ciencias, CUICBAS,
Universidad de Colima, Bernal D\'{\i}az del Castillo 340, Colima,
Colima, Mexico} \ead{paolo.amore@gmail.com}

\address{\ddag\ INIFTA (UNLP, CCT La Plata-CONICET), Divisi\'{o}n Qu\'{i}mica Te\'{o}rica,
Diag. 113 y 64 (S/N), Sucursal 4, Casilla de Correo 16, 1900 La
Plata, Argentina} \ead{fernande@quimica.unlp.edu.ar}

\begin{abstract}
We study both analytically and numerically the spectrum of inhomogeneous
strings with $\mathcal{PT}$-symmetric density. We discuss an exactly
solvable model of $\mathcal{PT}$-symmetric string which is isospectral to
the uniform string; for more general strings, we calculate exactly the sum
rules $Z(p) \equiv \sum_{n=1}^\infty 1/E_n^p$, with $p=1,2,\dots$ and find
explicit expressions which can be used to obtain bounds on the lowest
eigenvalue. A detailed numerical calculation is carried out for two
non-solvable models depending on a parameter, obtaining precise estimates of
the critical values where pair of real eigenvalues become complex.
\end{abstract}

\maketitle


\section{Introduction}

\label{sec:intro}

In the last years there has been great interest in the mathematical
properties of a class of non-hermitian operators with PT symmetry (for a
review see\cite{B07} and references therein). A good deal of this research
is based on a wide variety of simple models. In particular it is of great
interest to determine the conditions of unbroken symmetry under which the
eigenvalues are real. This unbroken symmetry takes place for a range of
values of a Hamiltonian parameter that in general increases with the quantum
number.

The purpose of this paper is the investigation of a new class of
PT-symmetric models: inhomogeneous vibrating strings. In a series of papers
Amore studied the spectral problems of inhomogeneous strings and drums\cite
{A10a,A10b,A11,A12,A13a,A13b}. In this paper we enlarge the class of such
problems to include vibrating strings with complex densities $\Sigma (x)$
that satisfy $\Sigma (-x)^{*}=\Sigma (x)$.

The paper is organized as follows: in section \ref{sec:PT_Strings}
introduce the problem, in section \ref{sec:PTSym} we discuss
$\mathcal{PT}$ symmetry, in section \ref{sec:RR} we discuss the
application of the Rayleigh-Ritz method to the study of
$\mathcal{PT}$ symmetric strings, in section \ref{sec:exact} we
introduce a family of $\mathcal{PT}$ symmetric strings, which
includes a string isospectral to the uniform string; in sections
\ref{sec:example1} and \ref{sec:example2} we discuss two examples
of $\mathcal{PT}$ symmetric strings which display a mixed
spectrum; finally in section \ref{sec:concl} we draw conclusions.

\section{PT-symmetric strings}
\label{sec:PT_Strings}

In this paper we consider the problem an inhomogeneous vibrating
string with density $\Sigma (x)$
\begin{equation}
-\frac{d^{2}}{dx^{2}}\psi _{n}(x)=E_{n}\Sigma (x)\psi _{n}(x),\,n=1,2,\ldots
\label{eq:Helmholtz1}
\end{equation}
and Dirichlet boundary conditions at the string ends $\psi (\pm
1/2)=0$. This equation can be straightforwardly converted
into\cite{A10a,A10b,A11,A12,A13a,A13b}
\begin{equation}
\frac{1}{\sqrt{\Sigma (x)}}\left[ -\frac{d^{2}}{dx^{2}}\right] \frac{1}{%
\sqrt{\Sigma (x)}}\phi _{n}(x)=E_{n}\phi _{n}(x)\ ,  \label{eq:Helmholtz2}
\end{equation}
where $\phi _{n}(x)\equiv \sqrt{\Sigma (x)}\psi _{n}(x)$. If $\Sigma (x)$ is
real positive function on $|x|\leq 1/2$, it follows that the operator $\hat{O%
}=\frac{1}{\sqrt{\Sigma (x)}}\left( -\frac{d^{2}}{dx^{2}}\right) \frac{1}{%
\sqrt{\Sigma (x)}}$ is hermitian. Another advantage of this form of the
eigenvalue equation is that the inverse operator $\hat{O}^{-1}=\sqrt{\Sigma
(x)}\left( -\frac{d^{2}}{dx^{2}}\right) ^{-1}\sqrt{\Sigma (x)}$ can be
directly expressed in terms of the Green's functions of the homogeneous
problem\cite{A13a,A13b}. In what follows we assume that $\Sigma (x)$ is
complex and $\mathcal{PT}$ symmetric.

In particular it is straightforward to generalize the results of \cite{A13a,A13b},
where exact expressions for the sum rules of inhomogeneous strings and drums
have been derived, to the present case. For instance, being $E_n$ the eigenvalues
of a $\mathcal{PT}$ symmetric string obeying Dirichlet boundary conditions at its ends,
we are interested in obtaining explicit expressions for the sum rules
\begin{equation}
Z_{DD}(s)=\sum_{n=1}^{\infty }E_{n}^{-s},\,s>0  \label{eq:Z_DD}
\end{equation}
with $s=1,2,\dots$. Analogous expressions should also be considered for the case
of different boundary conditions, as done in \cite{A13a}.

The case corresponding to $s=1$ can be directly obtained from
equation (11) of ref.~\cite{A13b} and reads~\footnote{We decompose
an arbitrary $\Sigma(x)$ in even and odd parts, $\Sigma(x) =
(\Sigma(x)+\Sigma(-x))/2 +  (\Sigma(x)-\Sigma(-x))/2$ and then use
the $\mathcal{PT}$-symmetry to establish that the even part of
$\Sigma(x)$ is real, whereas the odd part is imaginary.} \beq
Z_{DD}(1) &=& \int_{-a/2}^{+a/2} \left(\frac{a}{4} -
\frac{x^2}{a}\right) \ \Re \Sigma(x) dx .\eeq Therefore the
spectral sum rule $Z_{DD}(1)$ only depends upon the real part of
the density.

\section{$\mathcal{PT}$ symmetry}
\label{sec:PTSym}

$\mathcal{PT}$ symmetry is related to the antiunitary operator $\hat{P}\hat{T%
}$, where $\hat{P}$ and $\hat{T}$ are the parity and inversion operators,
respectively\cite{B07}. In general an antiunitary operator $\hat{A}$
satisfies\cite{W60}
\begin{eqnarray}
\hat{A}\left( f+g\right) &=&\hat{A}f+\hat{A}g  \nonumber \\
\hat{A}cf &=&c^{*}\hat{A}f,  \label{eq:antiunitary}
\end{eqnarray}
for any pair of vectors $f$ and $g$ and arbitrary complex number $c$, where
the asterisk denotes complex conjugation. In particular, $\hat{A}=\hat{P}%
\hat{T}$ satisfies the additional condition $\hat{A}^{2}=\hat{1}$.

In order to discuss the $\mathcal{PT}$ symmetry of inhomogeneous strings we
rewrite equation (\ref{eq:Helmholtz1}) as
\begin{equation}
\hat{L}\psi =-\left[ \frac{d^{2}}{dx^{2}}+E\Sigma (x)\right] \psi =0.
\end{equation}
It is clear that
\begin{equation}
\hat{A}\hat{L}\hat{A}^{-1}\hat{A}\psi =-\left[ \frac{d^{2}}{dx^{2}}%
+E^{*}\Sigma (x)\right] \hat{A}\psi =0,
\end{equation}
provided that $\Sigma (-x)^{*}=\Sigma (x)$ as already assumed above. We
appreciate that the eigenvalues are either real or pair of complex conjugate
numbers. In the former case we have
\begin{equation}
\hat{L}\hat{A}\psi =0.  \label{eq:brok_sym}
\end{equation}
One-dimensional eigenvalue equations with Dirichlet boundary conditions $%
\psi (\pm 1/2)=0$ do not exhibit degeneracy and (\ref{eq:brok_sym}) holds
only if $\hat{A}\psi =\lambda \psi $, from which it follows that $\hat{A}%
^{2}\psi =|\lambda |^{2}\psi =\psi $. In particular, when $\lambda
=\pm 1$ it follows from $\hat{A}\psi (x)=\psi (-x)^{*}=\pm \psi
(x)$ that the real and imaginary parts of $\psi (x)$ have definite
parity: $\Re \psi (-x)=\pm \Re \psi (-x)$, $\Im \psi (-x)=\mp \Im
\psi (-x)$. On the other hand, when symmetry is broken the
eigenfunctions for the pair of complex conjugate eigenvalues $E$
and $E^{*}$ are $\psi $ and $\hat{A}\psi $, respectively.


\section{Rayleigh-Ritz method}

\label{sec:RR}


In order to solve equation (\ref{eq:Helmholtz1}) we expand the solution as
\begin{equation}
\psi (x)=\sum_{m=1}^{\infty }c_{m}u_{m}(x),
\end{equation}
where
\begin{equation}
u_{m}(x)=\sqrt{2}\sin \left[ m\pi (x+1/2)\right] .
\end{equation}
Thus, the differential equation becomes the infinite matrix equation
\begin{equation}
\mathbf{LC}=\mathbf{0}
\end{equation}
where $\mathbf{C}$ is a column vector of the coefficients $c_{n}$ and $%
\mathbf{L}$ is a square matrix with elements
\begin{equation}
L_{mn}=n^{2}\pi ^{2}\delta _{mn}-E\Sigma _{mn},\,\Sigma
_{mn}=\int_{-1/2}^{1/2}u_{m}(x)\Sigma (x)u_{n}(x)\,dx
\end{equation}
The eigenvalues $E_{n}$ are given by the roots of
\begin{equation}
F=\det \mathbf{L}=0.  \label{eq:secular}
\end{equation}
In practice we truncate the matrices at dimension $N$ and calculate the
roots of equation (\ref{eq:secular}) for increasing values of $N$ till we
get the desired accuracy.

In all the cases discussed in this paper we have $F(E,\alpha )=0$, where $%
\alpha $ is a parameter in the string density. The critical values of $%
\alpha $ are given by $d\alpha /dE=0$ and we can obtain them from the set of
equations
\begin{equation}
\left\{ F(E,\alpha )=0,\,\partial F(E,\alpha )/\partial E=0\right\} .
\end{equation}
This strategy proved suitable for the treatment of parameter-dependent $%
\mathcal{PT}$-symmetric Hamiltonian operators\cite{FG13}.

\section{A class of solvable $\mathcal{PT}$-symmetric strings}
\label{sec:exact}

In the case of a string with Dirichlet boundary conditions at $\pm L$ Amore%
\cite{A11} showed that if the density satisfies the differential equation
\begin{equation}
4\Sigma ^{\prime \prime }(x)\Sigma (x)-5\Sigma ^{\prime }(x)^{2}-16\kappa
\Sigma (x)^{3}=0,  \label{eq:ODE}
\end{equation}
where $\kappa $ is an arbitrary constant, the solution is of the form
\begin{eqnarray}
\phi _{n}(x)=\sqrt{\frac{2}{\sigma (L)}}\Sigma (x)^{1/4}\sin \frac{n\pi
\sigma (x)}{\sigma (L)},  \label{eigen}
\end{eqnarray}
and
\[
\sigma (x)\equiv \int_{-L}^{x}\sqrt{\Sigma (y)}dy\ .
\]

The general solution to equation (\ref{eq:ODE}) for $L=1/2$ is
\[
\Sigma (x)=\frac{256c_{1}^{2}}{\left( c_{1}^{2}\left( c_{2}+x\right)
^{2}-256\kappa \right) {}^{2}},
\]
where $c_{1,2}$ are constants of integration. This solution contains the
\textsl{Borg string}\cite{B46}, an inhomogeneous string isospectral to the
homogeneous string, as a special case\cite{A10a}:
\[
c_{1}=\frac{2+\alpha }{8\alpha }\ \ \ ,\ \ \ c_{2}=\frac{(1+\alpha )^{2}}{%
\alpha ^{4}},
\]
where $\alpha >-1$ is an arbitrary parameter. In this case the density is
\[
\Sigma (x)=\frac{16(\alpha +1)^{2}}{(2\alpha x+\alpha +2)^{4}}\ .
\]

Remarkably, the spectrum of the Borg string is independent of $\alpha $ and
coincides with the spectrum of a homogeneous string ($\alpha =0$) of unit
length:
\[
E_{n}=n^{2}\pi ^{2}\ .
\]

The eigenfunctions are
\[
\phi _{n}(x)=\frac{2\sqrt{2}\sqrt{\alpha +1}}{2\alpha x+\alpha +2}\ \sin
\left( \frac{\pi (\alpha +1)n(2x+1)}{2\alpha x+\alpha +2}\right) \ .
\]

By means of a different choice of the constants of integration, for instance
$c_{1}=1$ and $c_{2}=i$, we obtain a complex density
\begin{equation}
\Sigma ^{(\mathcal{PT})}(x)=\frac{256}{\left( 256\kappa -x^{2}-2ix+1\right)
^{2}},  \label{sigmapt}
\end{equation}
that is invariant under the $\mathcal{PT}$ transformation.

In particular, the special case
\begin{equation}
\Sigma ^{(\mathcal{PT})}(x)=\frac{\left( \alpha ^{2}+64\right) ^{2}}{%
16(\alpha x+4i)^{4}}  \label{sigmapt2}
\end{equation}
is the $\mathcal{PT}$-symmetric analogous of the Borg string.

\begin{figure}[tbp]
\begin{center}
\bigskip\bigskip\bigskip \includegraphics[width=6cm]{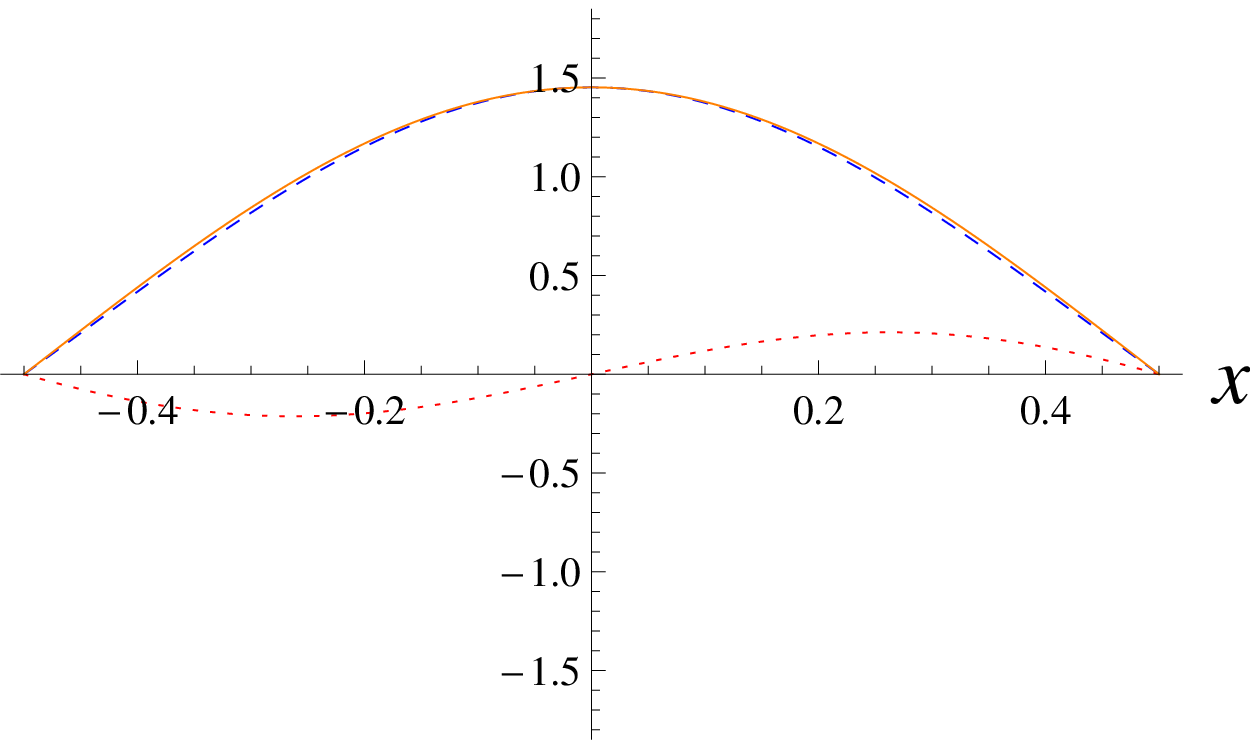} %
\includegraphics[width=6cm]{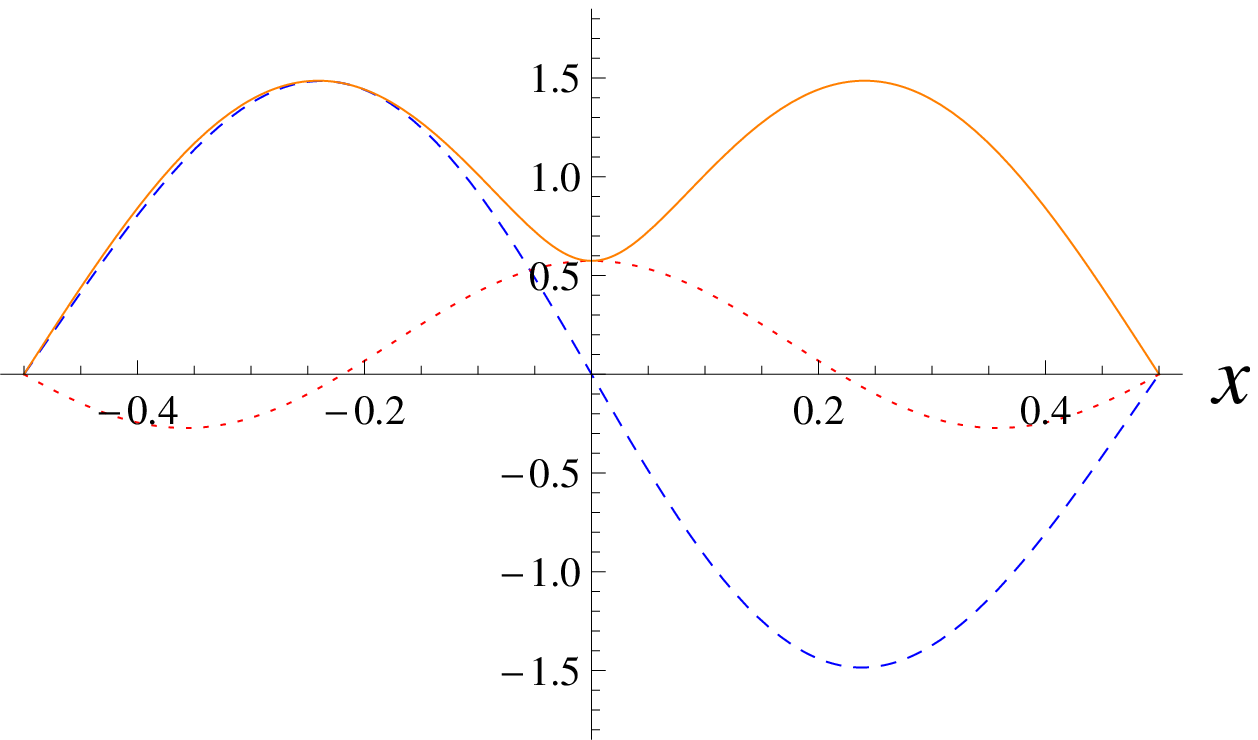} %
\includegraphics[width=6cm]{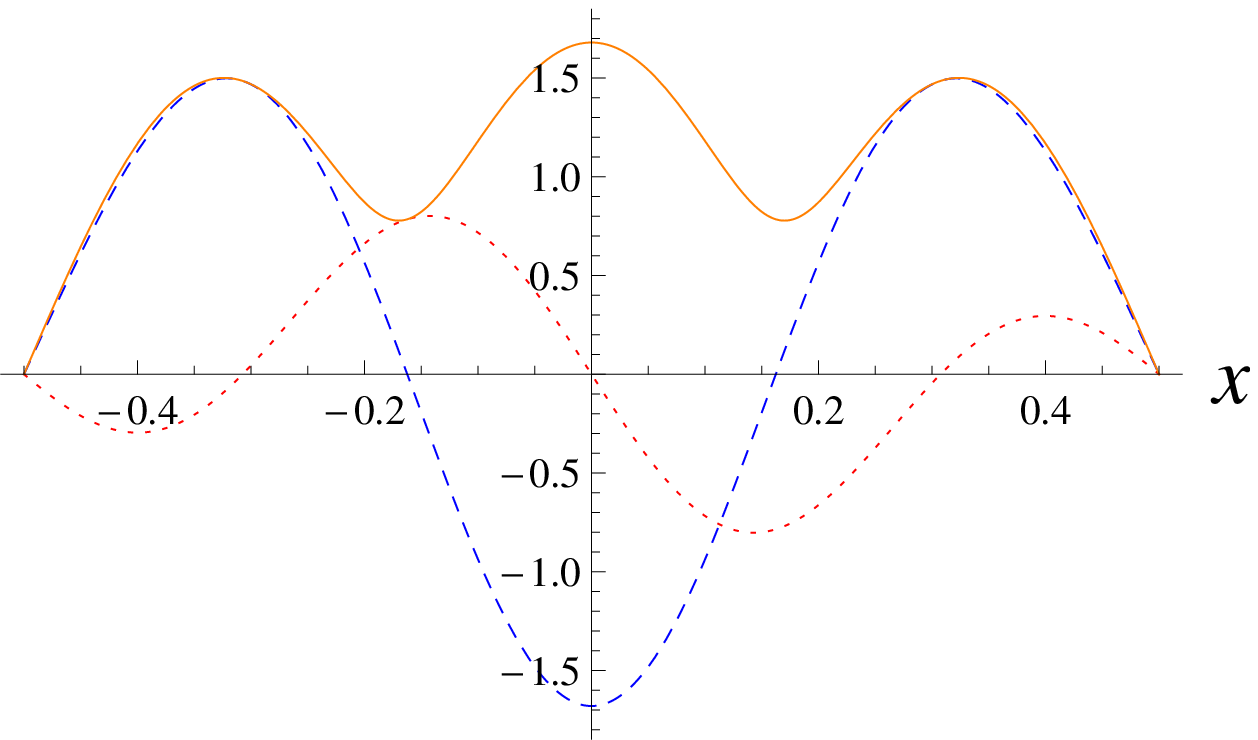} %
\includegraphics[width=6cm]{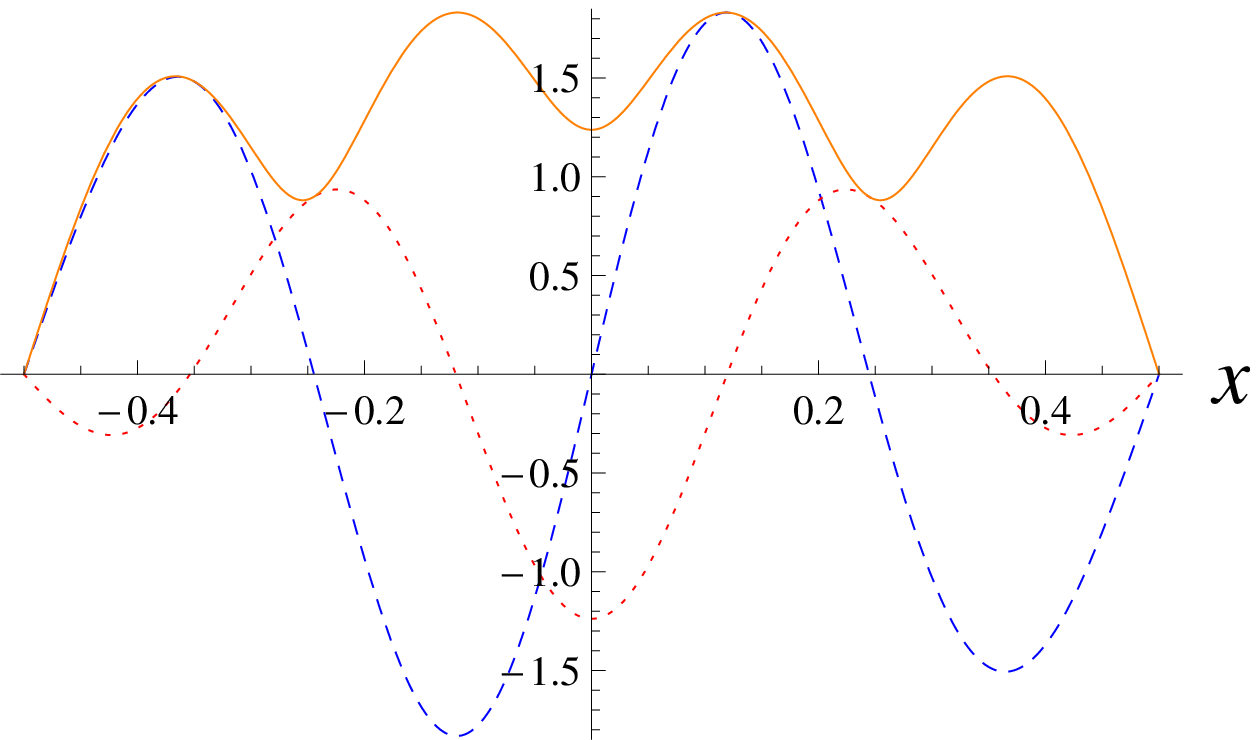}
\end{center}
\caption{First four eigenfunctions of the $\mathcal{PT}$-symmetric Borg
string for $\alpha=1$. The dashed and dotted lines are the real and
imaginary parts respectively; the solid line is the modulus of the
eigenfunction.}
\label{Fig_1}
\end{figure}

Using equation (\ref{eigen}) we obtain the eigenfunctions of the $\mathcal{PT%
}$-symmetric Borg string as
\begin{eqnarray}
\phi_n(x) &=& \sqrt{\frac{1}{2\sigma(L)}} \sqrt{\frac{\alpha ^2+64}{\alpha
^2 x^2+16}} \sqrt[4]{e^{-4 i \arg (\alpha x+4 i)}} \ \sin \frac{n \pi
\sigma(x)}{\sigma(L)}  \label{BorgPT}
\end{eqnarray}
where
\begin{eqnarray}
\sigma(x) = -\frac{\left(\alpha ^2+64\right) \sqrt{e^{-4 i \arg (\alpha x+4
i)}}}{4 \alpha (\alpha x-4 i)}-\frac{(\alpha -8 i) e^{\frac{1}{2} i \arg
\left(\frac{(\alpha +8 i)^2}{(\alpha -8 i)^2}\right)}}{2 \alpha }
\end{eqnarray}
and $n=1,2,\dots$.

Direct substitution of equation (\ref{BorgPT}) inside equation (\ref
{eq:Helmholtz2}) shows that these are indeed the exact eigenfunctions of a
string with density given in equation (\ref{sigmapt2}). The eigenvalues are
easily obtained
\begin{eqnarray}
E_n = \frac{1}{\phi_n(x)} \hat{O} \phi_n(x) = n^2 \pi^2
\end{eqnarray}

\textsl{Thus we see that this string has a real spectrum and that it is
isospectral to a homogeneous string with unit density}; on the basis of this
result we may conclude that one cannot ``hear'' the density of a $\mathcal{PT%
}$-symmetric string, if only Dirichlet boundary conditions are imposed, as
for the case of a real string.

Having the exact eigenfunctions at our disposal we may easily check that
these are orthogonal with respect to the operation
\begin{eqnarray}
\int_{-L}^{+L} \phi_n(x) \phi_{m}(x) dx = \pm \int_{-L}^{+L}
\phi_n^\star(-x) \phi_{m}(x) dx = \delta_{nm} \ .
\end{eqnarray}

Moreover
\begin{eqnarray}
\overline{\delta}(x,y) \equiv \sum_{n=1}^\infty \phi_n(x) \phi_n(y) =
\sum_{n=1}^\infty \phi^\star_n(-x) \phi^\star_n(-y)
\end{eqnarray}
has the Dirac-delta like properties
\begin{eqnarray}
\int_{-L}^L \overline{\delta}(x,y) \phi_m(x) dx &=& \phi_m(y) \\
\int_{-L}^L \overline{\delta}(x,y) \phi^\star_m(-x) dx &=& \phi^\star_m(-y)
\end{eqnarray}

\begin{figure}[tbp]
\begin{center}
\bigskip\bigskip\bigskip
\includegraphics[width=6cm]{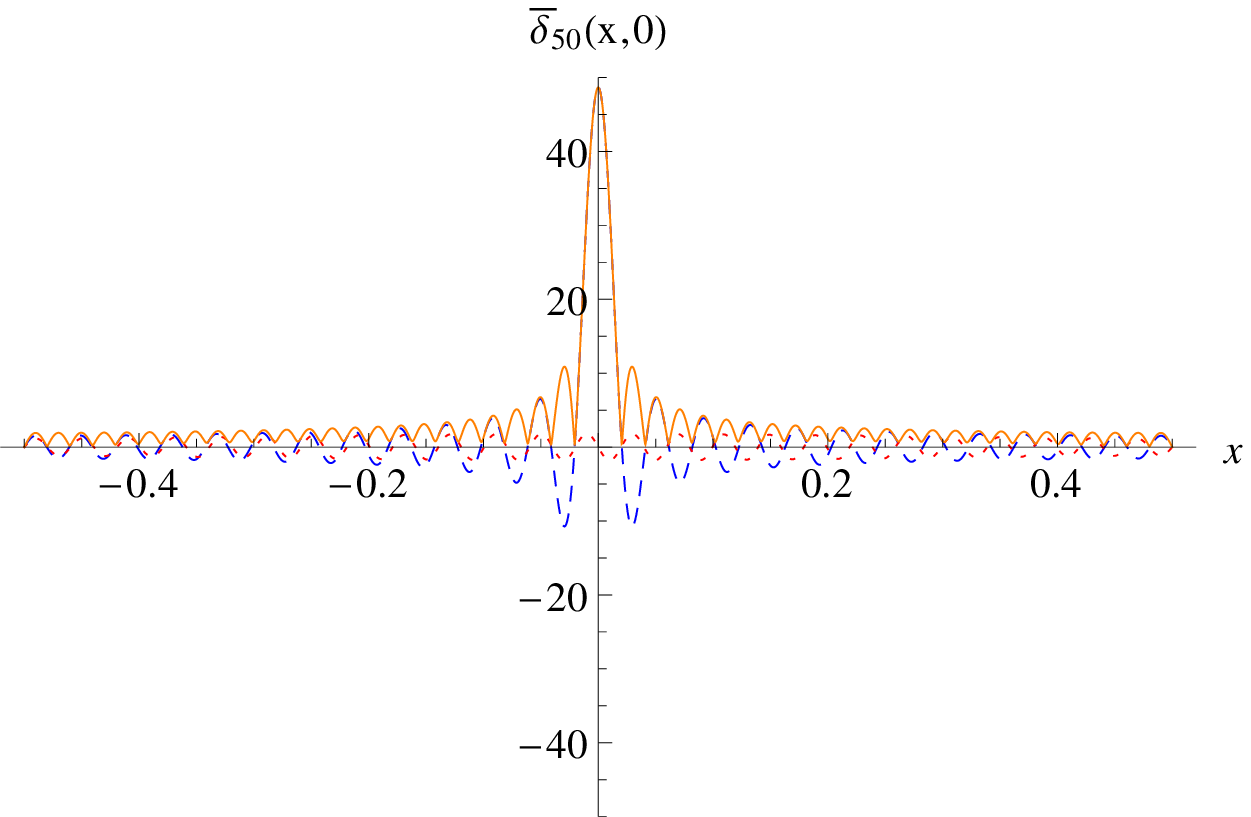} %
\includegraphics[width=6cm]{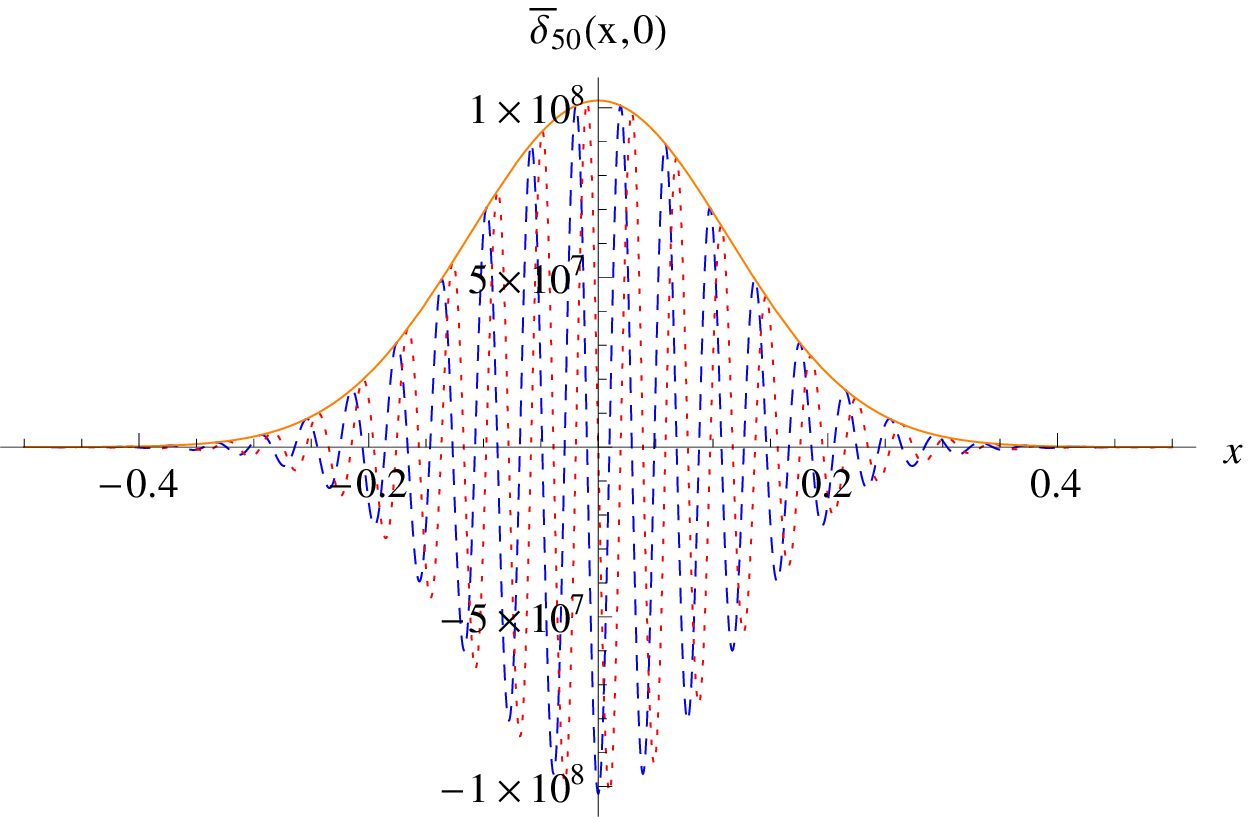} %
\end{center}
\caption{Approximate $\mathcal{PT}$-delta function $\overline{\delta}_{50}(x,0)$
for $\alpha=1/10$ (left plot) and $\alpha=1$ (right plot). The dashed and dotted lines are the real and
imaginary parts respectively; the solid line is the modulus.}
\label{Fig_delta}
\end{figure}

In Figure \ref{Fig_delta} we plot the approximation to
$\overline{\delta}(x,0)$ obtained restricting the sum to the first
$50$ terms, $\overline{\delta}_{50}(x,0)$, for $\alpha=1/10$ and
$\alpha=1$ (left and right plots respectively). Notice that for
$\alpha=0$ $\overline{\delta}(x,y)$ reduces to the Dirac delta
function and the imaginary part vanishes identically.

All the sum rules (\ref{eq:Z_DD}), $s=1,2,\ldots ,9$  calculated
analytically by means of the formulas given in reference \cite{A13a} agree
with the straightforward sums coming from the spectrum $E_{n}=n^{2}\pi ^{2}$.

This $\mathcal{PT}$-symmetric model is another example like the
Hamiltonian $\hat{H}=\hat{p}^{2}+\hat{x}^{2}+i\alpha \hat{x}$ with real
spectrum $E_{n}(\alpha )=(2n+1)+\alpha ^{2}/4$ for all real $\alpha $. We
can also add $\hat{H}=\hat{p}^{2}+i\alpha \hat{p}$ with the boundary
conditions $\psi (\pm L/2)=0$ with spectrum $E_{n}(\alpha )=n^{2}\pi
^{2}/L^{2}+\alpha ^{2}/4$.

\section{First example}
\label{sec:example1}

We consider a string with unit length ($L=1/2$) with density
\begin{equation}
\Sigma (x)=1+i\alpha x\ .  \label{eq:density_linear}
\end{equation}
Here we assume that $\alpha $ is a real arbitrary parameter. To begin with,
note that if the parameter-dependent string density $\Sigma (\alpha ,x)$
satisfies $\Sigma (-\alpha ,-x)=\Sigma (\alpha ,x)$ then the eigenvalues $%
E_{n}(\alpha )$ are symmetric about $\alpha =0$: $E_{n}(-\alpha
)=E_{n}(\alpha )$. This is exactly the case of the $\mathcal{PT}$-symmetric
density (\ref{eq:density_linear}).

In this case we use the exact formulas of Ref.\cite{A13a} and obtain:
\begin{eqnarray}
Z_{DD}(1) &=& \frac{1}{6} \\
Z_{DD}(2) &=& \frac{1}{90}-\frac{\alpha ^2}{5040} \\
Z_{DD}(3) &=& \frac{1}{945}-\frac{\alpha ^2}{30240} \\
Z_{DD}(4) &=& \frac{197 \alpha ^4}{10897286400}-\frac{17 \alpha ^2}{3742200}+%
\frac{1}{9450} \\
Z_{DD}(5) &=& \frac{17 \alpha ^4}{3269185920}-\frac{59 \alpha^2}{102162060}+%
\frac{1}{93555} \\
Z_{DD}(6) &=& -\frac{2341 \alpha^6}{1364608498176000}+\frac{16771 \alpha^4}{%
16672848192000}  \nonumber \\
&-& \frac{359 \alpha^2}{5108103000}+\frac{691}{638512875} \\
Z_{DD}(7) &=& -\frac{15773 \alpha^6}{22808456326656000}+\frac{46667 \alpha^4%
}{285105704083200}  \nonumber \\
&-& \frac{1237 \alpha^2}{148864716000}+\frac{2}{18243225} \\
Z_{DD}(8) &=& \frac{8458133 \alpha^8}{51894121836144107520000} -\frac{689371
\alpha^6}{3986227909984320000}  \nonumber \\
&+&\frac{736579 \alpha^4}{30490471131120000} - \frac{68197 \alpha^2}{%
70871446327500}+\frac{3617}{325641566250} \\
Z_{DD}(9) &=& \frac{111789019 \alpha^8}{1323300106821674741760000} -\frac{%
114268283 \alpha^6}{3308250267054186854400}  \nonumber \\
&+& \frac{5281763 \alpha^4}{1577881881035460000} -\frac{627073 \alpha^2}{%
5716963337085000}  \nonumber \\
&+& \frac{43867}{38979295480125}  \label{eq:ZDD_ex_1}
\end{eqnarray}

We may estimate the lowest eigenvalue of the string using the inequalities~%
\cite{B86}
\begin{eqnarray}
Z_{DD}(s)^{-1/s} \leq E_1^{(DD)} \leq \frac{Z(s)}{Z(s+1)}
\end{eqnarray}

Since the $Z_{DD}(n)$ are polynomials in $\alpha $, the occurrence of real
roots signals that $Z_{DD}(s)^{-1/s}$ can now take complex values, and
therefore that the spectrum cannot be completely real.

\begin{figure}[tbp]
\begin{center}
\bigskip\bigskip\bigskip \includegraphics[width=7cm]{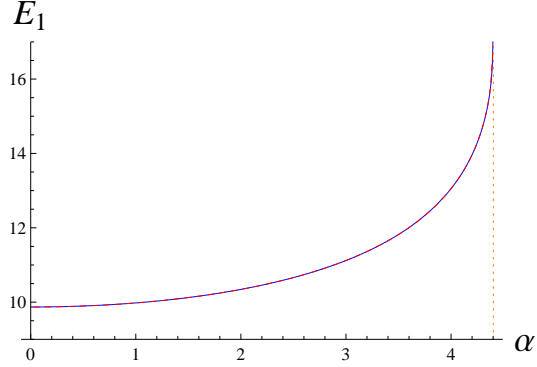}
\end{center}
\caption{Lowest eigenvalue as a function of $\alpha$ estimated using the sum
rules up to order $9$, and performing 3 repeated Shanks transformations. The
circles correspond to the numerical values obtained with collocation. }
\label{Fig_4}
\end{figure}

In ref.\cite{A13a} it has been proved that one can use the sequence of
approximations to the lowest eigenvalue $E_{1}\approx Z_{DD}(n)^{-1/n}$, to
obtain very accurate analytical approximations to $E_{1}$: using the same
strategy we have performed four repeated Shanks transformations obtaining a
precise analytical formula. This formula exhibits a singularity at $\alpha
^{\star }\approx 4.40272$ that is quite close to the accurate Rayleigh-Ritz
result $\alpha _{1}=4.397159356361900$. Figure \ref{Fig_4} shows the
estimate obtained with the Shanks transformations and the actual value of $%
\alpha _{1}$ (vertical line). We have also calculated the eigenvalues of the
$\mathcal{PT}$ string by means of a collocation method developed some time
ago\cite{ACF07}.

At $\alpha =0$ the eigenvalues of this string are those of the
homogeneous string. As $|\alpha |$ increases pairs of eigenvalues
start to approach each other and coalesce at a particular critical
value, $\alpha _{n}$, beyond which they become pairs of complex
conjugate numbers. More precisely, pairs of eigenvalues
$(E_{2n-1},E_{2n})$, $n=1,2,\ldots $, coalesce at the critical
point $\alpha _{n}$ where $E_{2n-1}(\alpha _{n})=E_{2n}(\alpha
_{n})=e_{n}$. It is most interesting that in this case $\alpha
_{1}>\alpha _{2}>\dots $ so that for each value of $\alpha
_{n+1}<\alpha <\alpha _{n}$ there is a finite number of real
eigenvalues!. This behaviour is completely
different from the one that takes place in a class of $\mathcal{PT}$%
-symmetric Hamiltonian operators, where $\alpha _{1}<\alpha _{2}<\dots $
\cite{B07}.

By means of the Rayleigh-Ritz method outlined in section \ref{sec:RR} we
calculated several pairs of critical parameters $\{e_{n},\alpha _{n}\}$ and
carried out nonlinear regressions of the form
\begin{equation}
\alpha _{n}=b+c|e_{n}|^{-s}.  \label{eq:nonlin_fit}
\end{equation}
For this particular string we obtained
\begin{eqnarray}
b &=&3.4685067\pm 0.00090795610  \nonumber \\
c &=&4.2281164\pm 0.027739157  \nonumber \\
s &=&0.53669526\pm 0.0023416105,
\end{eqnarray}
which suggests that there is an infinite number of real eigenvalues when $%
0<\alpha <b$.


\section{A $\mathcal{PT}$-string with real negative eigenvalues}

\label{sec:example2}


Another most interesting $\mathcal{PT}$ string is given by the density
\begin{equation}
\Sigma (x)=(1+i\alpha x)^{2}\ ,  \label{sigma_pt2}
\end{equation}
where $\alpha $ is a real parameter and $|x|\leq 1/2$ as before.

Once again we use the exact formulas of reference \cite{A13a} and obtain the
first $7$ sum rules:
\begin{eqnarray}
Z_{DD}(1) &=&\frac{1}{6}-\frac{\alpha ^{2}}{120} \\
Z_{DD}(2) &=&\frac{\alpha ^{4}}{50400}-\frac{\alpha ^{2}}{630}+\frac{1}{90}
\\
Z_{DD}(3) &=&-\frac{29\alpha ^{6}}{432432000}+\frac{\alpha ^{4}}{92400}-%
\frac{\alpha ^{2}}{4200}+\frac{1}{945} \\
Z_{DD}(4) &=&\frac{251\alpha ^{8}}{1029188160000}-\frac{23\alpha ^{6}}{%
378378000}+\frac{1499\alpha ^{4}}{567567000}  \nonumber \\
&-&\frac{\alpha ^{2}}{31185}+\frac{1}{9450} \\
Z_{DD}(5) &=&-\frac{3221\alpha ^{10}}{3519823507200000}+\frac{773\alpha ^{8}%
}{2514159648000}-\frac{3313\alpha ^{6}}{154378224000}  \nonumber \\
&+&\frac{83\alpha ^{4}}{170270100}-\frac{691\alpha ^{2}}{170270100}+\frac{1}{%
93555} \\
Z_{DD}(6) &=&\frac{16965349\alpha ^{12}}{4862213796375936000000}-\frac{%
759931\alpha ^{10}}{519467285937600000}  \nonumber \\
&+&\frac{1646627\alpha ^{8}}{11292767085600000}-\frac{1204631\alpha ^{6}}{%
230988417660000}+\frac{15047\alpha ^{4}}{192972780000}  \nonumber \\
&-&\frac{\alpha ^{2}}{2027025}+\frac{691}{638512875} \\
Z_{DD}(7) &=&-\frac{5405503\alpha ^{14}}{402869143128291840000000}+\frac{%
211469\alpha ^{12}}{31572816859584000000}  \nonumber \\
&-&\frac{460458127\alpha ^{10}}{520951478183136000000}+\frac{3219703\alpha
^{8}}{71559268981200000}-\frac{7523137\alpha ^{6}}{7259635983600000}
\nonumber \\
&+&\frac{565843\alpha ^{4}}{49497518070000}-\frac{3617\alpha ^{2}}{%
62026965000}+\frac{2}{18243225}
\end{eqnarray}
The fact that $Z_{DD}(n)$ can take negative values signals that
part of the spectrum must be complex.

\begin{figure}[tbp]
\begin{center}
\bigskip\bigskip\bigskip \includegraphics[width=7cm]{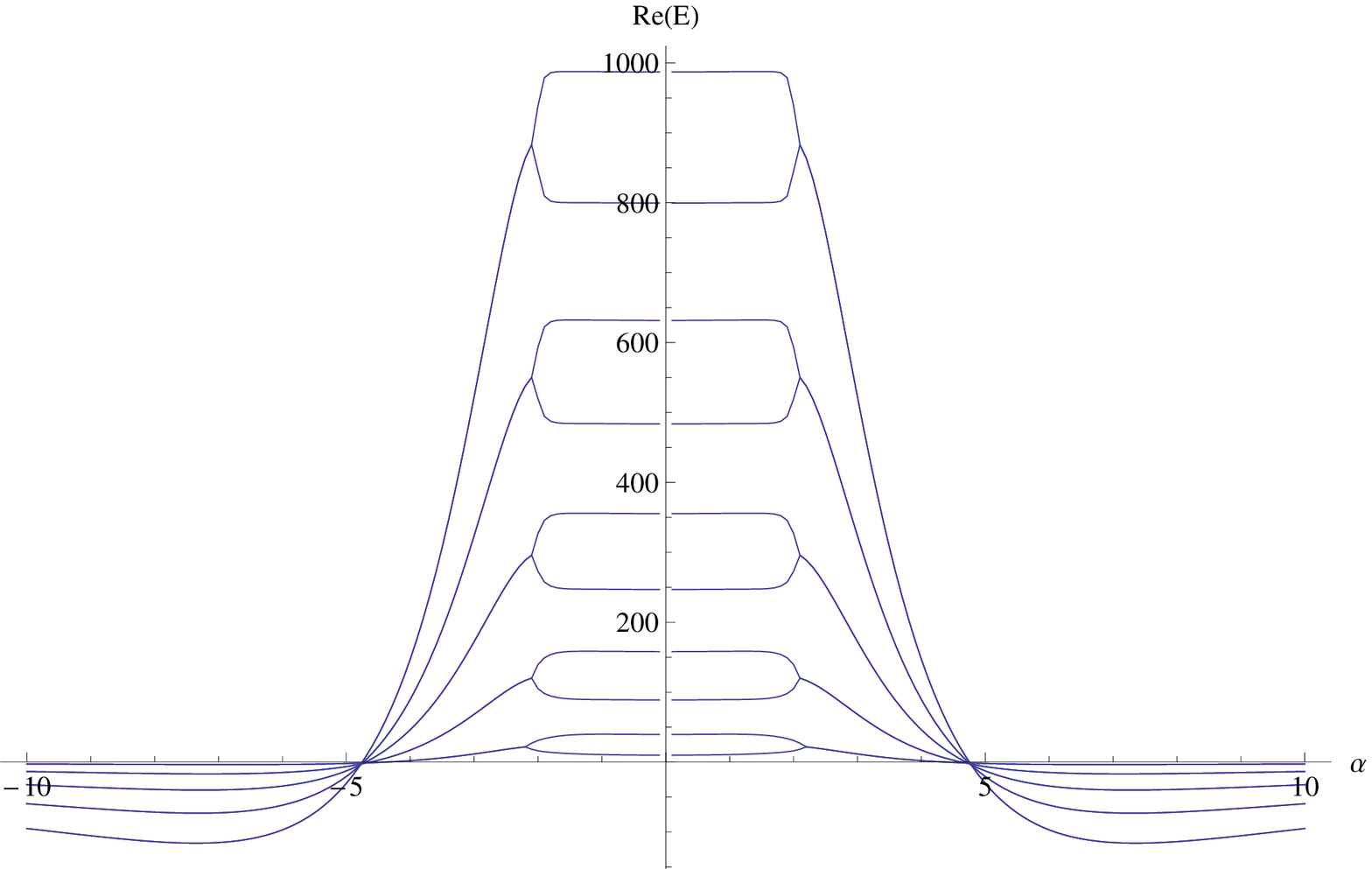} %
\includegraphics[width=7cm]{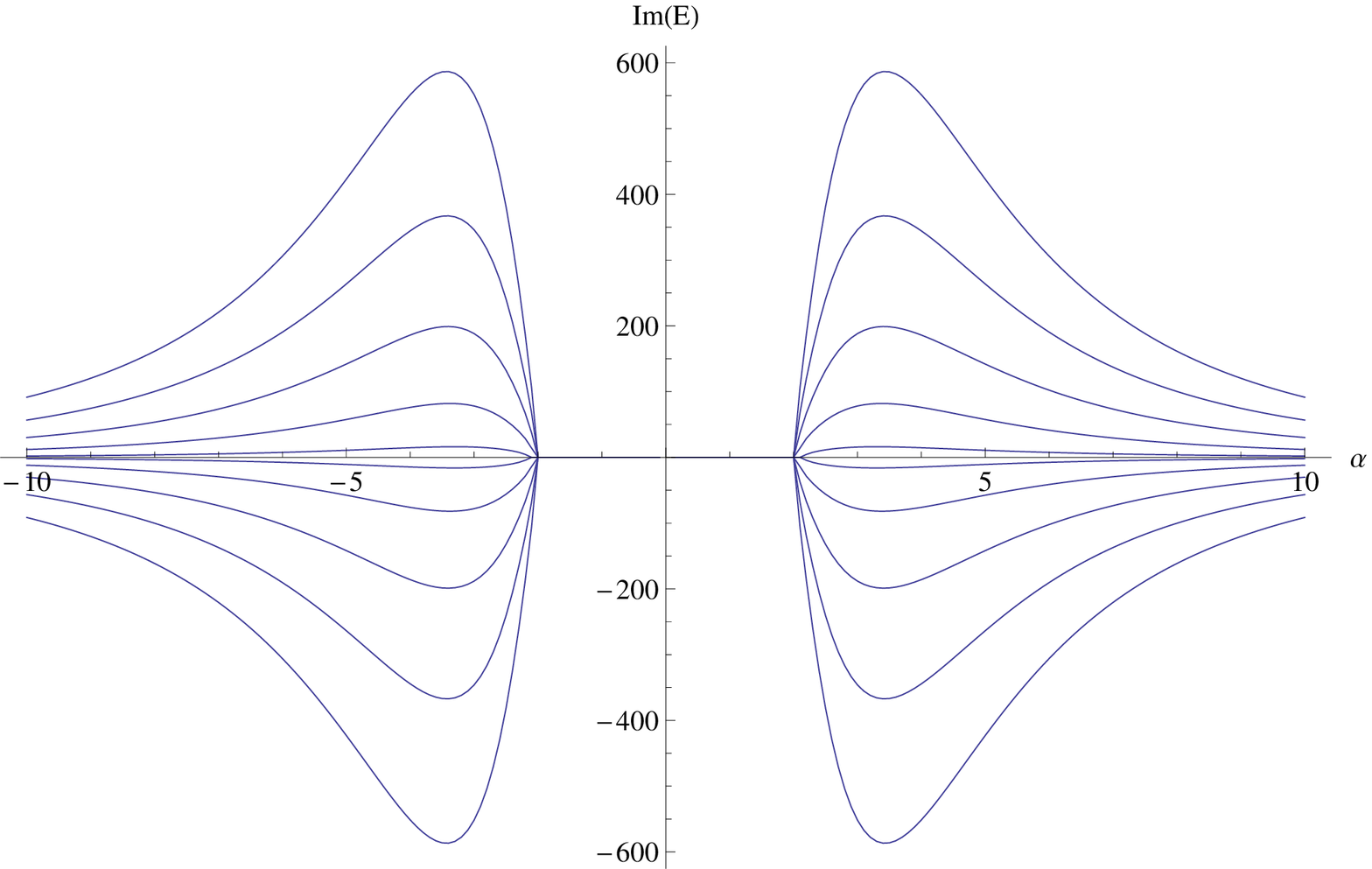}
\end{center}
\caption{Real and imaginary parts of the first eight eigenvalues
of the string with density (\ref{sigma_pt2}) for $-10 \leq \alpha
\leq 10$.} \label{ptsym2a}
\end{figure}

\begin{figure}[tbp]
\begin{center}
\bigskip\bigskip\bigskip \includegraphics[width=7cm]{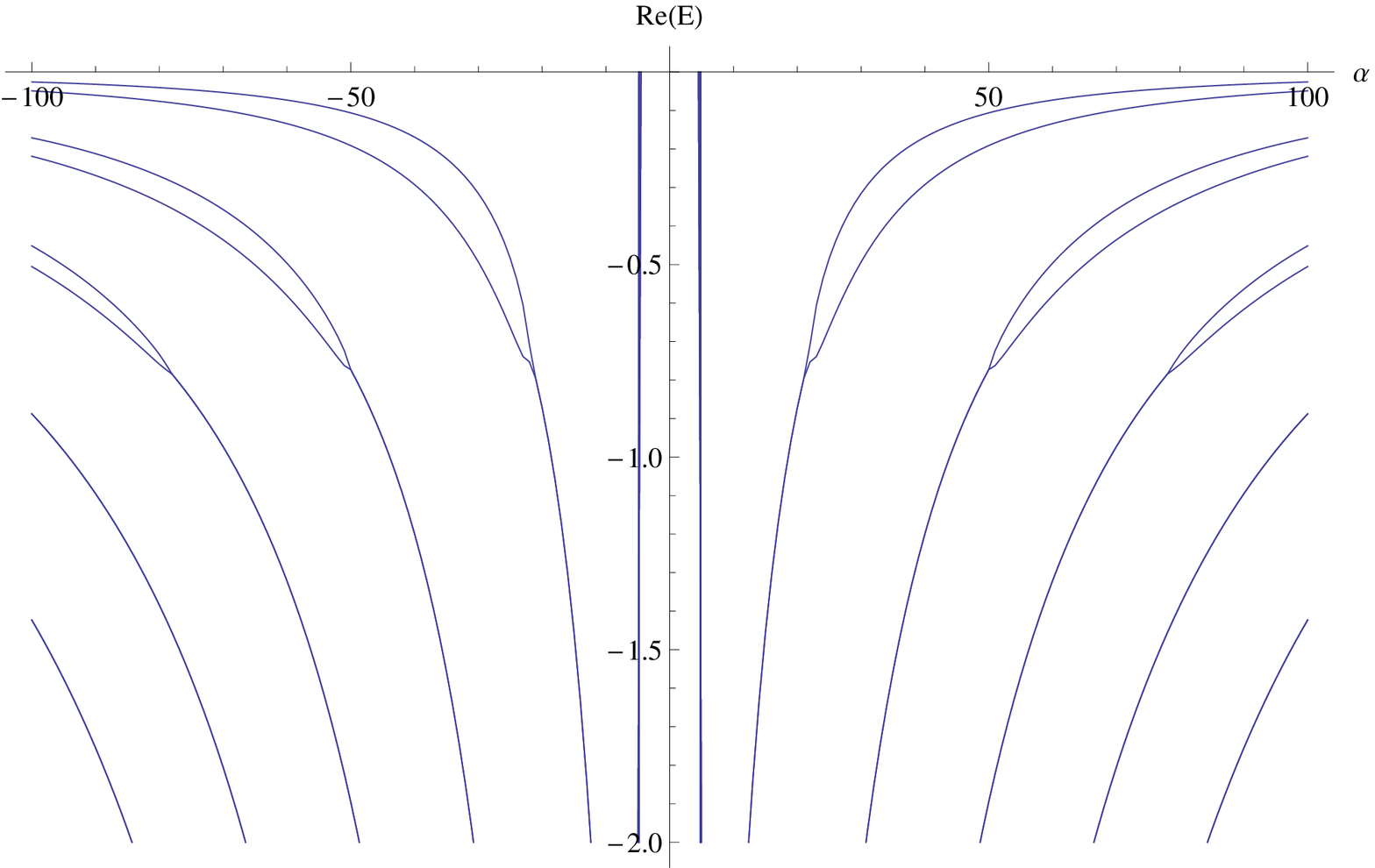} %
\includegraphics[width=7cm]{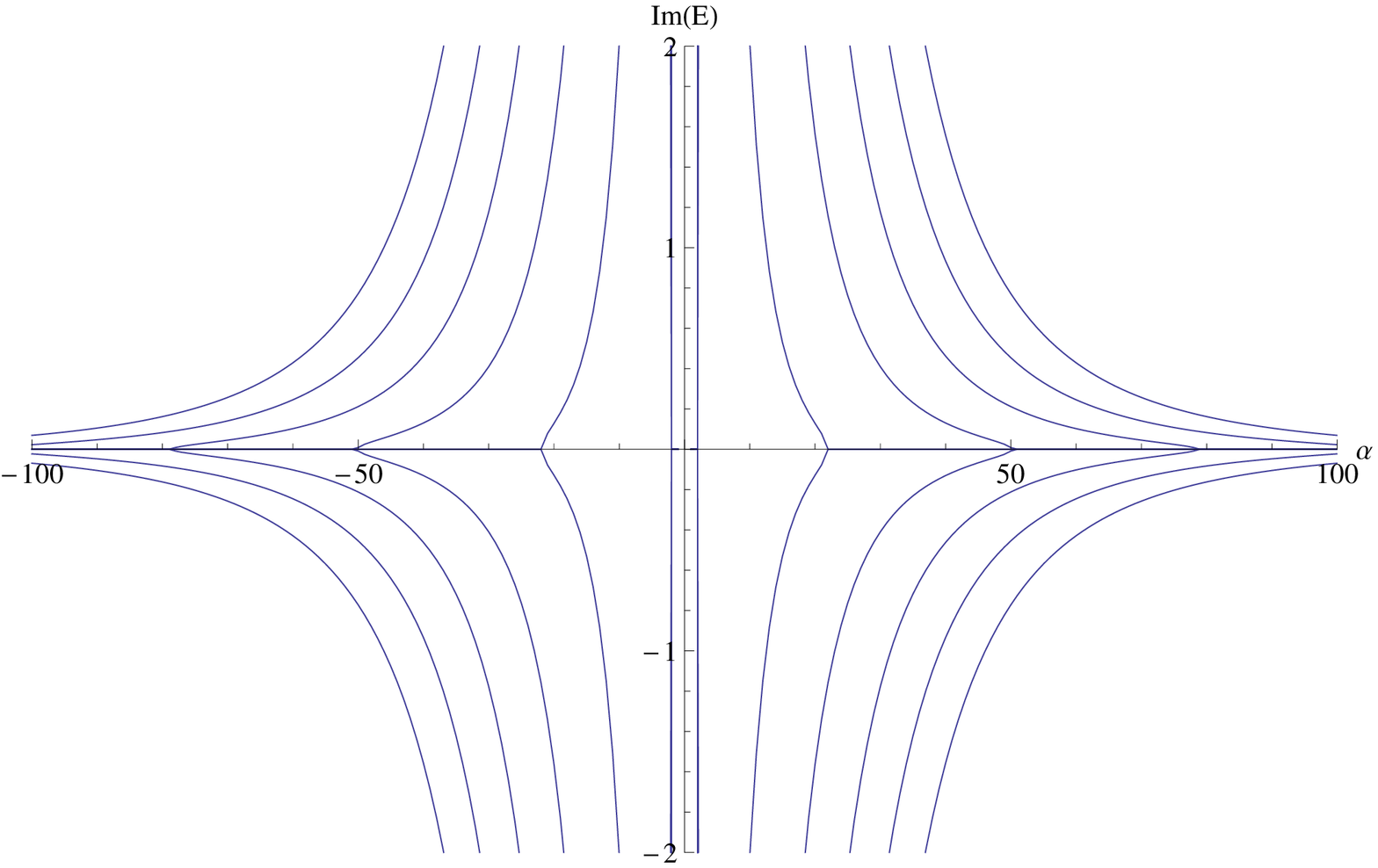}
\end{center}
\caption{Real and imaginary parts of the first eight eigenvalues
of the string with density (\ref{sigma_pt2}) for $-100 \leq \alpha
\leq 100$.} \label{ptsym2b}
\end{figure}

A useful strategy to obtain approximate solutions to the string with density
(\ref{sigma_pt2}) is to apply the Rayleigh-Ritz method as indicated in
section \ref{sec:RR} or the collocation approach to the operator $\hat{O}$.
In Figure \ref{ptsym2a} we show the numerical results for the real and
imaginary parts of the first eight eigenvalues of the string with density (%
\ref{sigma_pt2}) for $-10\leq \alpha \leq 10$: these results are obtained
using a collocation approach with a grid with $2000$ points\cite{ACF07}.
Looking at the right plot we see that the eigenvalues are real when $%
-2\lesssim \alpha \lesssim 2$. In Figure \ref{ptsym2b} we show the same
results for $-100\leq \alpha \leq 100$: in this case pairs of real \textsl{%
negative} eigenvalues appear when $\alpha $ reaches the critical values. The
first pair coalesce at $\pm \alpha _{1}$, where $\alpha _{1}=21.90376732248$.

\begin{figure}[tbp]
\begin{center}
\bigskip\bigskip\bigskip \includegraphics[width=7cm]{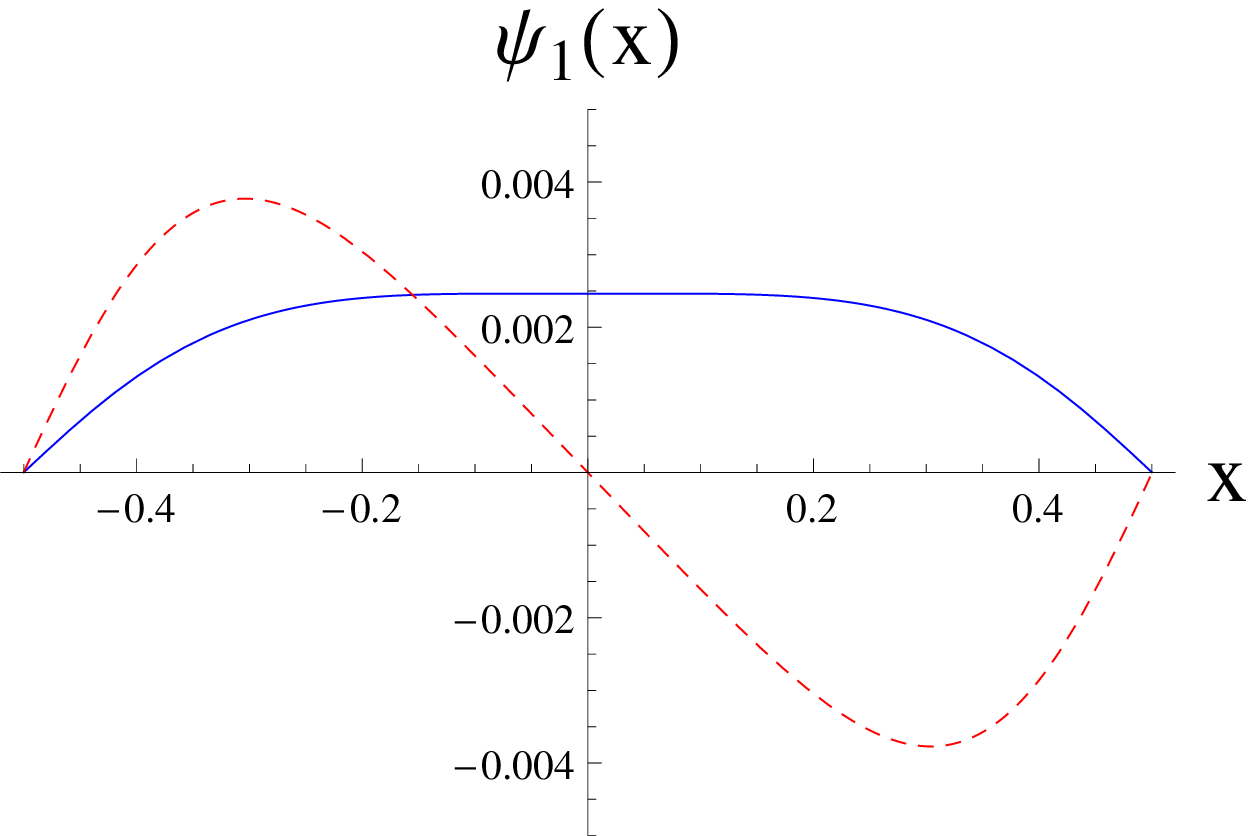} %
\includegraphics[width=7cm]{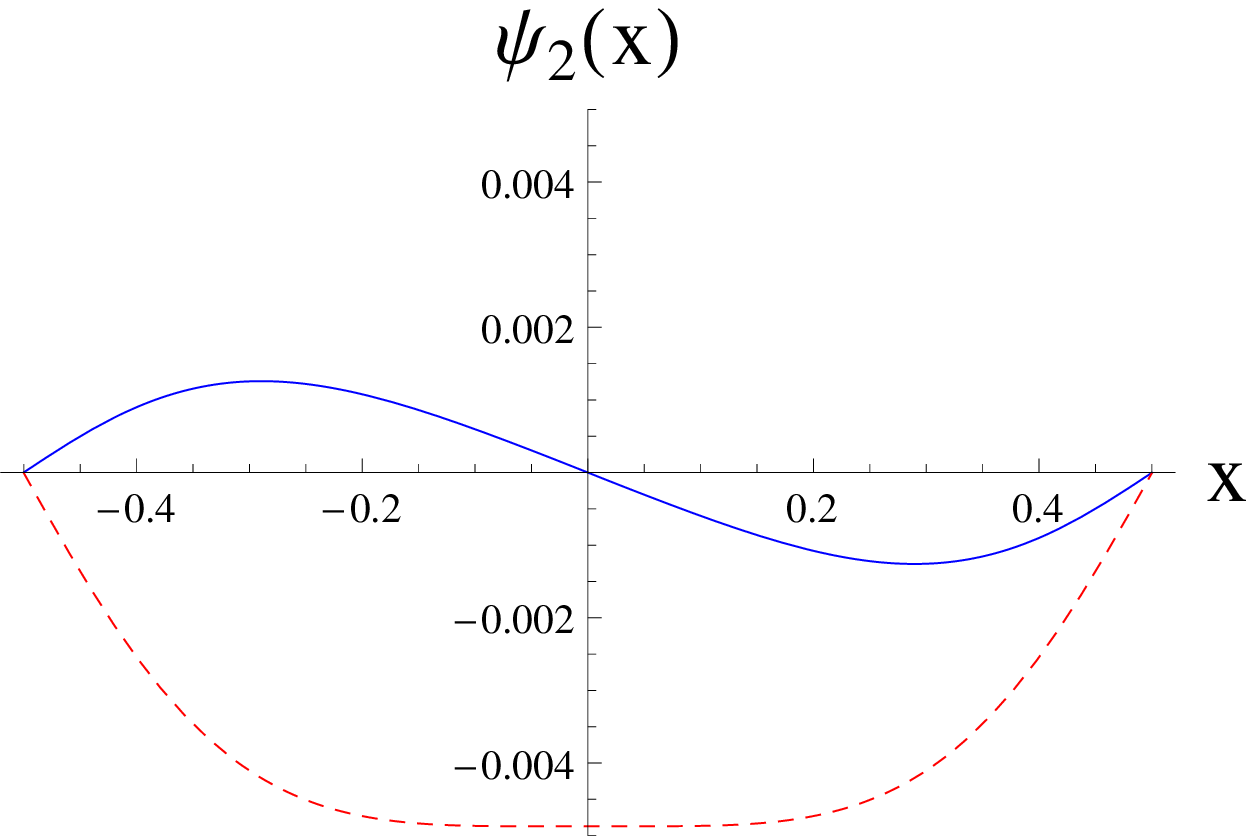}
\end{center}
\caption{Real (solid) and imaginary (dashed) parts of the eigenfunctions of
the first two modes of the string (\ref{sigma_pt2}) for $\alpha=30$.}
\label{ptsym2wfa}
\end{figure}

\begin{figure}[tbp]
\begin{center}
\bigskip\bigskip\bigskip \includegraphics[width=7cm]{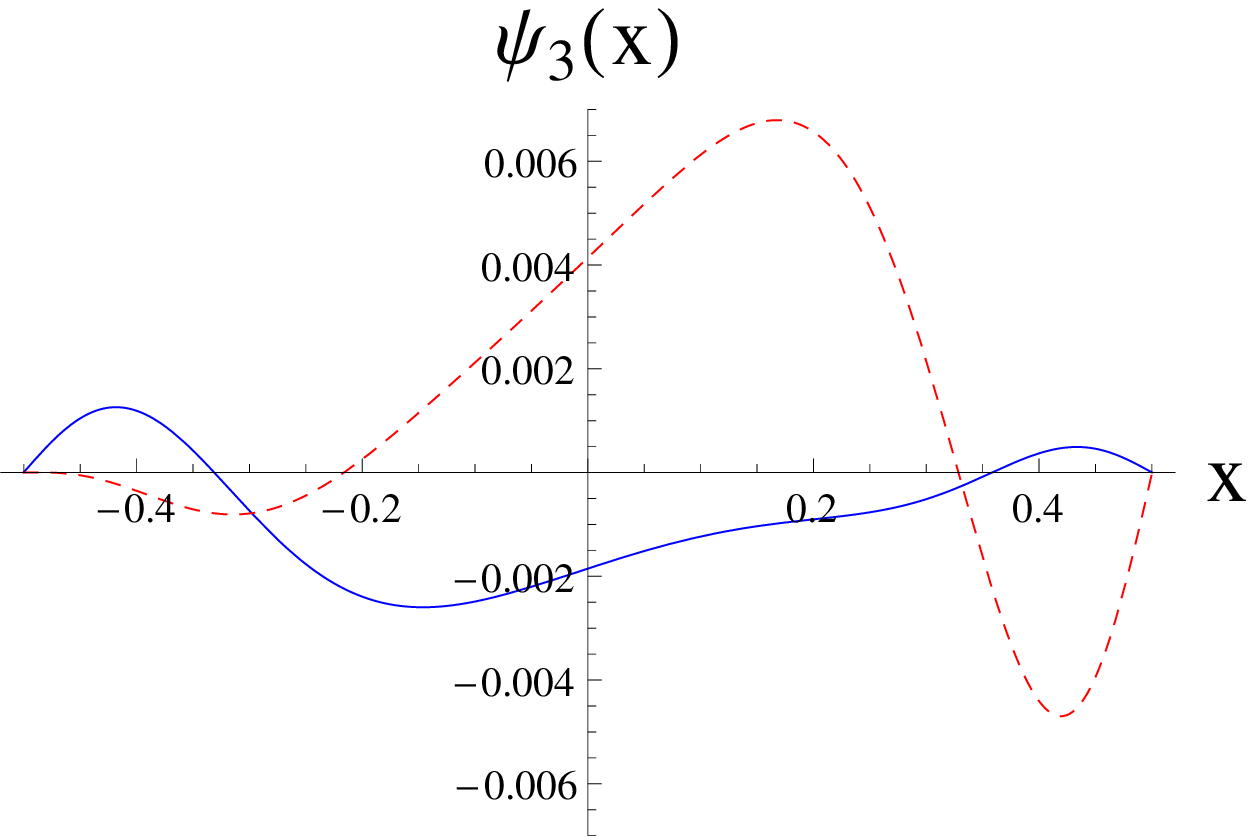} %
\includegraphics[width=7cm]{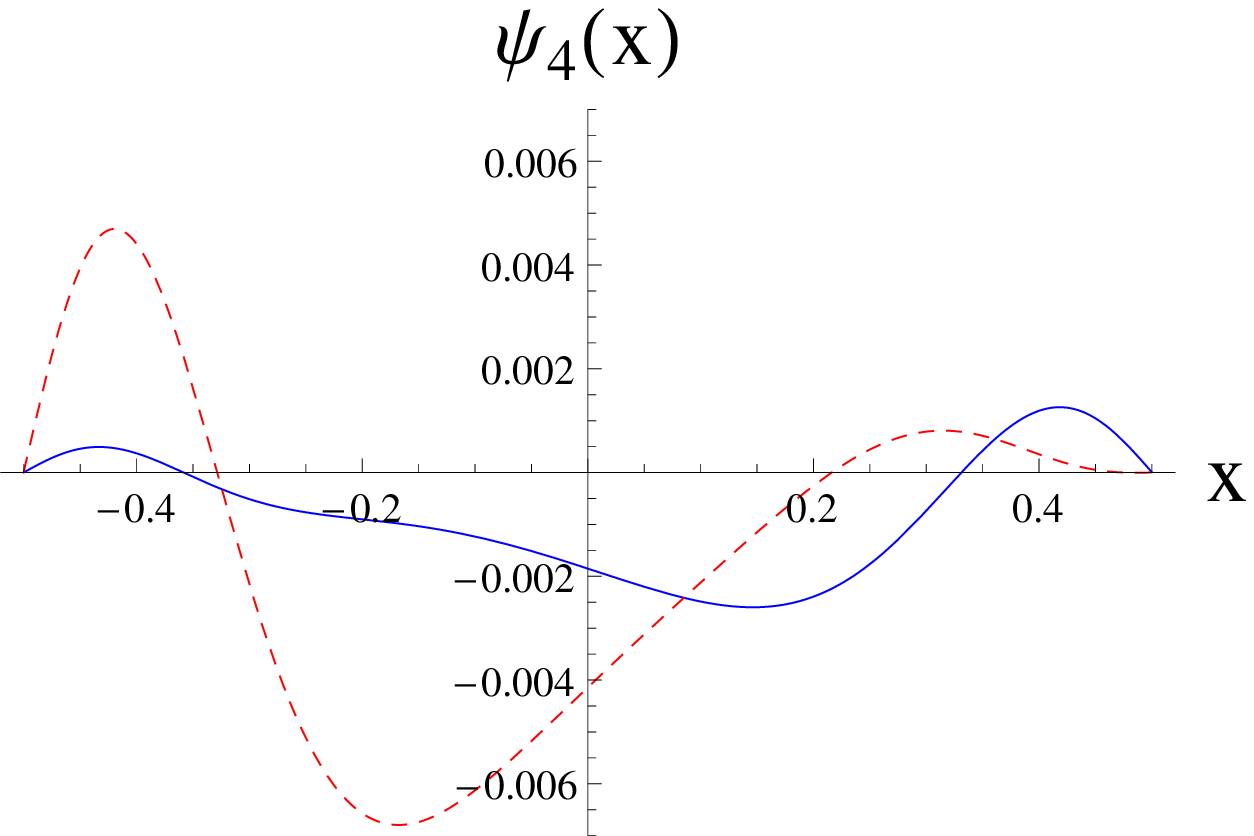}
\end{center}
\caption{Real (solid) and imaginary (dashed) parts of the eigenfunctions of
the third and fourth modes of the string (\ref{sigma_pt2}) for $\alpha=30$.}
\label{ptsym2wfb}
\end{figure}

It is interesting to focus on the second region, where the
spectrum contains pairs of \textsl{real negative eigenvalues}. In
particular, we choose $\alpha =30$, where a single pair of such
eigenvalues appears. In figure \ref {ptsym2wfa} we plot the real
and imaginary parts of the eigenfunctions of the first two modes,
whose energies are real and negative. In figure \ref {ptsym2wfb}
we plot the real and imaginary parts of the eigenfunctions of the
third and fourth modes, which exhibit complex conjugate
eigenvalues. These solutions are numerical approximations to the
eigenfunctions of equation(\ref{eq:Helmholtz1}). We may get an
idea of the precision of our collocation calculation from the
results of Table \ref{table1}, which compares the numerical sum
rules for this string at $\alpha =30$ with the exact ones shown
above. It follows from those figures that $\psi _{1}(-x)^{*}=\psi
_{1}(x)$, $\psi _{2}(-x)^{*}=-\psi _{2}(x)$ and that $\psi
_{3}(-x)^{*}=\psi _{4}(x)$ in complete agreement with the
discussion at the end of section~\ref {sec:PTSym}.

In this case the nonlinear fitting yields two sets of critical parameters
\begin{eqnarray}
b &=&-0.77692697\pm 2.7920949\times 10^{-5}  \nonumber \\
c &=&13.397511\pm 0.29472502  \nonumber \\
s &=&1.8798088\pm 0.0072380433,
\end{eqnarray}
for $e_{n}<0$ and

\begin{eqnarray}
b &=&2.0000002\pm 1.1418782\times 10^{-5}  \nonumber \\
c &=&0.70814609\pm 0.00032819029  \nonumber \\
s &=&0.50227919\pm 0.00015369406.
\end{eqnarray}
for $e_{n}>0$. In the latter case we conjecture that the exact asymptotic
relation may be
\begin{equation}
\alpha _{n}=2+\frac{1}{\sqrt{2e_{n}}}.
\end{equation}

\begin{table}[tbp]
\caption{Comparison between the sum rules for the string (\ref{sigma_pt2})
at $\alpha=30$ obtained using the numerical values obtained with collocation
on a grid with $N=2000$ and the exact sum rules. }
\label{table1}
\bigskip
\par
\begin{center}
\begin{tabular}{|l|r|c|c|}
\hline
$q$ & $Z^{(DD)}_{num}(q)$ & $Z^{(DD)}_{exact}(q)$ & $\left|
Z^{(DD)}_{num}(q)/Z^{(DD)}_{exact}(q)-1\right|$ \\ \hline
1 & $-7.32958160 + 4.13 \times 10^{-10} \ i$ & $-\frac{22}{3}$ & $0.00051$
\\
2 & $+14.65396825 - 2.58 \times 10^{-9} \ i$ & $\frac{4616}{315}$ & $2.1
\times 10^{-10}$ \\
3 & $-40.33560515 + 1.24 \times 10^{-8} \ i$ & $-\frac{5450752}{135135}$ & $%
4.74 \times 10^{-10}$ \\
4 & $+117.80838771 - 5.31 \times 10^{-8} \ i$ & $\frac{9472421696}{80405325}$
& $7.34 \times 10^{-10}$ \\
5 & $-353.88875146 + 2.12 \times 10^{-7} \ i$ & $-\frac{973145269792}{%
2749862115}$ & $1.01 \times 10^{-10}$ \\
6 & $+1082.41430676 - 8.11 \times 10^{-7} \ i$ & $\frac{5139579853771120064}{%
4748255660523375}$ & $1.29 \times 10^{-9}$ \\
7 & $-3351.2084737 + 3.01 \times 10^{-6} \ i$ & $-\frac{2636911102632544448}{%
786853795172445}$ & $1.56 \times 10^{-9}$ \\ \hline
\end{tabular}
\end{center}
\par
\bigskip\bigskip
\end{table}

\section{Conclusions}

\label{sec:concl}

The purpose of this paper is to enlarge the class of $\mathcal{PT}$%
-symmetric models with the addition of parameter-dependent inhomogeneous
strings with complex densities that satisfy  $\Sigma ^{\star }(\alpha
,-x)=\Sigma (\alpha ,x)$. We discussed an exactly solvable example with real
spectrum for all values of $\alpha $. This trivial inhomogeneous string is
isospectral with the homogeneous one (a $\mathcal{PT}$-symmetric analog of
the string found by Borg~\cite{B46} some time ago).

We also discussed two nontrivial strings for which one can obtain exact sum
rules thus extending Amore's result \cite{A13a} to the $\mathcal{PT}$%
-symmetric realm. The accurate calculation of the critical parameters
revealed that one of the strings exhibits real positive spectrum and the
other one both positive and negative eigenvalues. Obviously, such negative
eigenvalues cannot take place when the operator $\hat{O}$ is Hermitian.

Another interesting feature of the $\mathcal{PT}$-symmetric strings is that
the behaviour of the critical parameters is different from that one observed
in $\mathcal{PT}$-symmetric Hamiltonians like $\hat{H}=\hat{p}^{2}+\hat{x}%
^{4}+i\alpha \hat{x}$ or $\hat{H}=\hat{p}^{2}+i\hat{x}^{3}+i\alpha \hat{x}$%
\cite{B07} (and references therein).

\verb||\ack
The research of P.A. was supported by the Sistema Nacional de Investigadores
(M\'exico).

\verb||

\section*{References}

\verb||


\begin{thebibliography}{99}
\bibitem{B07}  Bender C M 2007 \textit{Rep. Prog. Phys.} \textbf{70} 947.

\bibitem{A10a}  Amore P 2010 \textit{Ann. Phys.} \textbf{325} 2679.

\bibitem{A10b}  Amore P 2010 \textit{J. Math. Phys.} \textbf{51} 052105.

\bibitem{A11}  Amore P 2011 \textit{Ann. Phys.} \textbf{326} 2315.

\bibitem{A12}  Amore P 2012 \textit{J. Math. Phys.} \textbf{53} 123519.

\bibitem{A13a}  Amore P 2013 Exact sum rules for inhomogeneous strings
arXiv:1302.4366 [math-ph]

\bibitem{A13b}  Amore P 2013 Exact sum rules for inhomogeneous drums
arXiv:1302.4371 [math-ph]

\bibitem{W60}  Wigner E 1960 \textit{J. Math. Phys.} \textbf{1} 409.

\bibitem{B46}  Borg G 1946 \textit{Acta Math.} \textbf{78} 1.

\bibitem{FG13}  F. M. Fern\'{a}ndez and J. Garcia, Critical parameters for
non-hermitian Hamiltonians, arXiv:1305.5164 [math-ph].

\bibitem{ACF07}  Amore P, Cervantes M, and Fern\'{a}ndez F M 2007 \textit{J.
Phys. A} \textbf{40} 13047.

\bibitem{B86}  Berry M V 1986 \textit{J. Phys. A} \textbf{19} 2281.
\end{thebibliography}
\end{document}